\documentclass[aps,floats]{revtex4}
\usepackage{amsmath,amssymb}
\usepackage{graphicx,epsfig}

\begin{document}
\bibliographystyle {plain}

\def\oppropto{\mathop{\propto}} 
\def\opsimeq{\mathop{\simeq}}
\def\opoverderline{\mathop{\overline}}
\def\operarrow{\mathop{\longrightarrow}}
\def\opsim{\mathop{\sim}}

\def\fig#1#2{\includegraphics[height=#1]{#2}}
\def\figx#1#2{\includegraphics[width=#1]{#2}}


\title{ Low-temperature dynamics of Long-Ranged Spin-Glasses : \\
 full hierarchy of relaxation times via real-space renormalization
  } 


 \author{ C\'ecile Monthus }
  \affiliation{Institut de Physique Th\'{e}orique,\\
  CNRS and CEA Saclay \\
 91191 Gif-sur-Yvette, France}

\begin{abstract}
We consider the long-ranged Ising spin-glass with random couplings decaying as a power-law of the distance, in the region of parameters where the spin-glass phase exists with a positive droplet exponent. For the Metropolis single-spin-flip dynamics near zero temperature, we construct  via real-space renormalization the full hierarchy of relaxation times of the master equation for any given realization of the random couplings. We then analyze the probability distribution of dynamical barriers as a function of the spatial scale. This  real-space renormalization procedure represents a simple explicit example of the droplet scaling theory, where the convergence towards local equilibrium on larger and larger scales is governed by a strong hierarchy of activated dynamical processes, with valleys within valleys.

\end{abstract}

\maketitle

\section{ Introduction}

The relaxation dynamics of disordered systems towards thermal equilibrium
when starting from a random initial condition at $t=0$ (see for instance \cite{bouchaud,bouchaudetal,youngbook,houches} and references therein) can be interpreted
by the following picture :  at time $t$, there exists some spatial coherence length $L(t)$ such that the smaller lengths $L < L(t)$ are in quasi-local-equilibrium,
whereas the bigger lengths $L > L(t)$ are still completely out of equilibrium.
In pure systems, these phenomena of phase ordering are well understood
 \cite{brayreview} and the coherence length grows algebraically $L_{pure}(t) \sim t^{1/z} $ with the dynamical exponent $z=2$ for non-conserved dynamics \cite{brayreview}. For pure systems, this domain growth is possible even at zero-temperature because domain walls can still diffuse and annihilate.
 In the presence of quenched disorder however, 
the dynamics requires thermal activation, since exactly
at zero temperature, the dynamics stops on the first encountered local minimum.
Within the droplet scaling theory
proposed both for spin-glasses \cite{heidelberg,Fis_Hus} 
and for directed polymers in random media \cite{Fis_Hus_DP}, the dynamical barriers $B(L)$
grow as a power law of the spatial length $L$
\begin{eqnarray}
B(L) \sim L^{\psi} u
\label{defpsi}
\end{eqnarray}
with some barrier exponent $\psi>0$, and where $u$ is a random variable of order $O(1)$.
The time $t(L)$ needed to go over this barrier by thermal activation then follows the activated form at small temperature $T=1/\beta$
\begin{eqnarray}
t(L) \simeq e^{ \beta B(L) } = e^{\beta L^{\psi} u }
\label{ttyp}
\end{eqnarray}
This means that the characteristic
length-scale $L(t)$ associated to time $t$ grows only logarithmically in time as
\begin{eqnarray}
L(t) \sim \left( \frac{ \ln t }{\beta}  \right)^{\frac{1}{\psi}}
\label{typltime}
\end{eqnarray}
In the field of spin-glasses, this logarithmic behavior has
 remained controversial, both in numerical studies (see \cite{SG_activ}
in favor of activated dynamics and \cite{SG_alge} in favor of power-law dynamics) and in experiments \cite{SG_expe}
 because the dynamics is very
slow both in Monte-Carlo simulations and in real life :
as a consequence, the maximal equilibrated length $L_{max} $ measured
 at the end of the dynamics is usually rather small,
so that various fits of the data are possible.
Besides the scaling of the growing length $L(t)$, many studies have also been devoted
 to more complicated temperature cycling experiments that display rejuvenation and memory \cite{bouchaud,bouchaudetal,youngbook,houches}. 
The important point for the present discussion 
is that these phenomena require some hierarchical organization 
of valleys within valleys, where the rejuvenation due to short length scales
does not destroy the memory of large length scales which are effectively frozen.

This expected hierarchical organization of valleys within valleys in space-time
strongly suggests that renormalization is the most appropriate tool to characterize the dynamics of disordered systems. In this direction, the real-space Strong Disorder Renormalization (see \cite{us_review} for a review) has given a lot of asymptotic exact results for the dynamics of many classical disordered models, including in particular random walks in random media
 \cite{sinairg,ferenc,sinaibiasdirectedtraprg,sinai_golosov,sinai_energy,rg_toy,greg,RWchannel,rw2d},
trap models \cite{traprg}, classical spin chains \cite{rfimrg}, reaction-diffusion \cite{readiffrg,pierre}, directed percolation
\cite{directedperco}, zero-range processes
\cite{zerorange}, exclusion processes 
\cite{exclusion}, contact processes
\cite{contact}, coupled oscillators
\cite{coupledoscillators}
and elastic networks
\cite{elasticnetworks}.
The Strong Disorder Renormalization has also 
been formulated {\rm in configuration space }
for the master equation of arbitrary disordered models \cite{us_rgconfig}. 
Another real-space renormalization procedure has also been introduced to study the largest relaxation time for pure and random ferromagnets in various geometries \cite{us_quantum}.

 In the present paper, we consider long-ranged spin-glasses and study the dynamics near zero temperature via a block renormalization in real space. This standard block renormalization is the simplest framework to take into account the hierarchy of the long-ranged couplings and to obtain simple renormalization rules. This procedure is thus different from the Strong Disorder Renormalization mentioned above, where the smallest barriers are decimated recursively. However, when dynamical barriers $B(L)$ grow as a power of the spatial length $L$ (Eq. \ref{defpsi}), we expect that the decimation of the smallest barriers or the decimation of the smallest lengths should be able to describe the same physics, i.e. more precisely should yield the same typical exponents (see \cite{us_fractal} where
the comparison between Strong Disorder renormalization and Block renormalization is discussed in detail for disorder quantum models governed by Infinite disordered Fixed Points).

The paper is organized as follows. The long-ranged spin-glass models are described
in section \ref{sec_model}, together with the Metropolis dynamics on which we 
focus. In section \ref{sec_rgstatic}, we recall the zero-temperature real-space renormalization to construct the two ground-states \cite{c_rgsg}. In section \ref{sec_rgrule}, we derive the basic renormalization rule for the Metropolis dynamics near zero temperature. In section \ref{sec_dynamics}, we describe how this basic renormalization rule can be used to construct explicitly the full hierarchy of relaxation times in each given disordered sample. Finally in section \ref{sec_barrier}, we analyze the properties
of the probability distributions of dynamical barriers. Our conclusions are summarized in section \ref{sec_conclusion}.

\section{ Models and notations }

\label{sec_model}

In real spin-glasses with RKKY interactions, the coupling $J(r)$ between two spins separated by a distance $r$ decays only as a power-law of the distance
and is of random sign
\begin{eqnarray}
J^{RKKY}(r) \simeq \pm \frac{1}{r^3}
\label{jrkky}
\end{eqnarray}
So besides the short-ranged spin-glasses that have been most studied, it is also important to better understand long-ranged spin-glasses. In the presence of power-law interactions, the dimensionality of the space is not such an essential parameter as for the short-ranged case, so that most studied have focused on the one-dimensional long-ranged spin-glass as we now recall.

\subsection{ One-dimensional Long-Ranged Spin-glass }

The one-dimensional Long-Ranged Spin-glass of $L$ classical spins $S_i=\pm 1$
is defined by the energy function
\begin{eqnarray}
 U(S_1,...,S_L)  = - \sum_{1 \leq i <j \leq L} J_{ij} S_i S_j 
\label{sgLR}
\end{eqnarray}
where the random couplings decay 
as a power-law of the distance $r=\vert j=i \vert$ with exponent $\sigma$
\begin{eqnarray}
J_{ij} && = \Delta(\vert j-i \vert)  \epsilon_{ij}
\nonumber \\
\Delta(r) && = \frac{1}{r^{\sigma}}
\label{defjijini}
\end{eqnarray}
The $\epsilon_{ij} $ 
are independent identical $O(1)$ random variables of zero mean.

The Gaussian distribution
 \begin{eqnarray}
L_{2}(\epsilon)  =  \frac{1}{\sqrt{4 \pi} } e^{- \frac{\epsilon^2}{4}}
\label{gaussian}
\end{eqnarray}
has been the most studied in the literature \cite{kotliar,BMY,KY,KYgeom,KKLH,KKLJH,Katz,KYalmeida,Yalmeida,KDYalmeida,LRmoore,KHY,KH,mori,wittmann,us_overlaptyp,us_dynamic,us_chaos}, but the L\'evy 
symmetric stable laws $L_{\mu}(\epsilon) $ of index $1<\mu \leq 2$ have also been considered \cite{c_rgsg}
 \begin{eqnarray}
L_{\mu}(\epsilon) && = \int_{-\infty}^{+\infty} \frac{dk}{2 \pi} e^{-i k \epsilon  - \vert k \vert^{\mu} }
\label{levy}
\end{eqnarray}
The Gaussian distribution of Eq. \ref{gaussian} corresponds to the particular case $\mu=2$. The other cases $1<\mu<2$ correspond to distributions with the following power-law tail
 \begin{eqnarray}
L_{1<\mu<2}(\epsilon) && \opsimeq_{ \epsilon  \to \pm \infty} \frac{A_{\mu}}{\vert \epsilon \vert^{1+\mu}}
\nonumber \\
A_{\mu} && \equiv \frac{\Gamma(1+\mu)}{\pi} \sin \left(\frac{\pi \mu}{2} \right)
\label{levytail}
\end{eqnarray}
In the field of spin-glasses, the case of L\'evy distributions
 of the couplings has been already studied for 
the mean-field fully connected geometry \cite{cizeau,janzen,mezard,neri,boettcher_levy} and for the nearest-neighbor model in dimension $d=3$ \cite{andresen}.

In this paper, we will only consider the region
\begin{eqnarray}
 \sigma  > \frac{1}{\mu}
\label{regionmu}
\end{eqnarray}
(in particular $\sigma > \frac{1}{2} $ for the Gaussian case $\mu=2$)
where the ground-state energy is extensive in the number of spins (see \cite{c_rgsg} for more details).

Within the droplet scaling theory \cite{heidelberg,Fis_Hus}, the most important property of the spin-glass phase
is the droplet exponent $\theta$ that governs the scaling 
of the renormalized random coupling $J_L$ with the length $L$
 \begin{eqnarray}
J_L \propto L^{\theta}
\label{deftheta}
\end{eqnarray}
Whereas in short-ranged spin-glasses,  the droplet exponent
 $\theta^{SR}(d)$ is non-trivial for dimensions $d>1$, 
the droplet exponent for Gaussian long-ranged spin-glasses 
is known exactly to be $\theta^{LR}(d,\sigma) = d-\sigma $
  \cite{Fis_Hus,BMY} in the region where
 $ \theta^{LR}(d,\sigma)> \theta^{SR}(d)$. In particular in dimension $d=1$,
where the short-ranged droplet exponent is known to be $\theta^{SR}(d=1) =-1$,
the droplet exponent of the
Gaussian long-ranged spin-glass is known exactly \cite{BMY,Fis_Hus}
\begin{eqnarray}
\theta^{LR}_{Gauss}(d=1,\sigma) && = 1-\sigma \ \ \ \ \ \ \ \ \ \ \ \ \ \ \ \ \ \ {\rm for } \ \ \frac{1}{2} < \sigma < 2
\nonumber \\
\theta^{LR}_{Gauss}(d=1,\sigma) 
&& = \theta^{SR}(d=1)=-1 \ \  {\rm for } \ \  2 \leq \sigma  
\label{thetaLRd1gauss}
\end{eqnarray}
Note however that the numerical measures via Monte-Carlo on sizes $L \leq 256$
(see Fig. 13 and Table III of \cite{KY}) are not a clear support of this theoretical expectation,
in particular in the region $\sigma \to (1/2)^+$ where the theoretical prediction of Eq. \ref{thetaLRd1gauss} corresponds to $\theta^{LR}(d=1,\sigma\to (1/2)^+) \to (1/2)^-$, whereas the numerical results of \cite{KY} display a saturation around $\theta \simeq 0.3$. The origin of this discrepancy has remained unclear over the years.
The interpretation proposed in  \cite{KY} is that
Eq. \ref{thetaLRd1gauss} is nevertheless exact in the whole region  
$\frac{1}{2} < \sigma < 2 $
as predicted by the theoretical derivations \cite{BMY,Fis_Hus},
and despite their numerical results \cite{KY}.
Another interpretation could be that the saturation seen in the numerics is
meaningful, and that Eq. \ref{thetaLRd1gauss} is valid only in the region 
$\frac{2}{3} < \sigma < 2 $.

For the L\'evy distribution of index $1<\mu<2$, the generalization reads \cite{c_rgsg}
\begin{eqnarray}
\theta^{LR}_{\mu}(d=1,\sigma) && = \frac{2}{\mu}-\sigma \ \ \ \ \ \ \ \ \ \ \ \ \ \ \ \ \ \ {\rm for } \ \ \frac{1}{\mu} < \sigma < \frac{2}{\mu}+1
\nonumber \\
\theta^{LR}_{\mu}(d=1,\sigma) && = \theta^{SR}(d=1)=-1 \ \  {\rm for } \ \   \sigma 
> \frac{2}{\mu}+1
\label{thetaLRd1levy}
\end{eqnarray}
In this paper, we will only consider the region of positive droplet exponent
\begin{eqnarray}
\theta^{LR}_{\mu}(d=1,\sigma) && = \frac{2}{\mu}-\sigma >0 
\ \ \ \ \  {\rm for } \ \ \frac{1}{\mu} < \sigma < \frac{2}{\mu}
\label{regionmutheta}
\end{eqnarray}
 where the spin-glass phase exists in a finite region of temperature
(i.e. the region $\frac{1}{2} < \sigma < 1$ for the Gaussian case).

\subsection{ Related Dyson hierarchical Spin-glass }

In the field of long ranged models, it is very useful to consider
their Dyson hierarchical analogs, where real space renormalization 
procedures are usually easier to define as a consequence of the hierarchical structure. The Dyson hierarchical ferromagnetic Ising model
 \cite{dyson} has been much studied by both mathematicians
\cite{bleher,gallavotti,book,jona} and physicists \cite{baker,mcguire,Kim,Kim77,us_dysonferrodyn}. More recently, various Dyson hierarchical versions of
disordered systems have been considered,
in particular Anderson localization models \cite{bovier,molchanov,krit,kuttruf,fyodorov,EBetOG,fyodorovbis,us_dysonloc}, random fields Ising models
\cite{randomfield,us_aval} and spin-glasses \cite{franz,castel_etal,castel_parisi,castel,angelini}.

The Dyson hierarchical spin-glass model of $L=2^N$ spins
is defined by the following recurrence for the energy function  \cite{castel_etal,castel_parisi,castel,angelini}
\begin{eqnarray}
U_{N}(S_1,S_2,...,S_{2^N}) &&  = U_{N-1}^{(a)}(S_1,S_2,...,S_{2^{N-1}})
+U_{N-1}^{(b)}(S_{2^{N-1}+1},S_{2^{N-1}+2},...,S_{2^{N}})
\nonumber \\
&&  -  \sum_{i=1}^{2^{N-1}} \sum_{j=2^{N-1}+1}^{2^N} J_{N-1}(i,j) S_i S_j 
\label{sgDyson}
\end{eqnarray}
(where the notation $U_{N-1}^{(a)} $ and $U_{N-1}^{(b)} $ means that
these two energies are two independent realizations for the two half-systems
before the introduction of the couplings of the second line).
The first terms for $N=1$ and $N=2$ reads
\begin{eqnarray}
U_{1}(S_1,S_2)  && = -J_0(1,2) S_1 S_2
\nonumber \\
U_{2}(S_1,S_2,S_3,S_4)  && = -J_0(1,2) S_1 S_2-J_0(3,4) S_3 S_4 \nonumber \\
&&
- J_1(1,3) S_1 S_3 - J_1(1,4) S_1 S_4
- J_1(2,3) S_2 S_3 - J_1(2,4) S_2 S_4
\label{sgDyson1}
\end{eqnarray}
At generation $n$, associated to the length scale $L_n=2^n$,
the couplings $J_n(i,j) $ read
\begin{eqnarray}
J_n(i,j)=\Delta_n \epsilon_{ij}
\label{jndysonsg}
\end{eqnarray}
where $ \epsilon_{ij}$ are independent random variables of zero mean
as in Eq. \ref{defjijini}, distributed with the Gaussian (Eq \ref{gaussian})
or the L\'evy law (Eq. \ref{levy}). At generation $n$,
the characteristic scale $\Delta_n$ is chosen to decay exponentially
with the number $n$ of generations,
in order to mimic the power-law decay of Eq. \ref{defjijini}
 with respect to the length scale $L_n=2^{n}$
\begin{eqnarray}
\Delta_n = 2^{-n \sigma} =  \frac{1}{L_n^{\sigma}} 
\label{deltandysonsg}
\end{eqnarray}
Then one expects that many scaling properties will be the same.
In particular, the condition for the extensivity of the energy is the same as Eq. \ref{regionmu}, and in the interesting region of positive droplet exponent
where the spin-glass phase exists in a finite region of temperature,
 the droplet exponent is given by the same formula as Eq. \ref{regionmutheta}
(see \cite{c_rgsg} for more details).

\subsection{ Master Equation for the Metropolis dynamics }

We consider the long-ranged spin-glass of $L=2^N$
spins of Eq. \ref{sgLR}  or of its Dyson analog of Eq. \ref{sgDyson}.
There are 
\begin{eqnarray}
{\cal N}= 2^L=2^{2^N}
\label{calN}
\end{eqnarray}
 possibles configurations 
${\cal C}=(S_1,S_2,...,S_L) $.
The stochastic relaxational dynamics
 towards the Boltzmann equilibrium 
\begin{eqnarray}
P_{eq}({\cal C}) = \frac{ e^{- \beta U({\cal C})} }{Z}
\end{eqnarray}
where $Z$ is the partition function
\begin{eqnarray}
Z = \sum_{\cal C}  e^{- \beta U({\cal C})}
\label{partition}
\end{eqnarray}
can be described by the master equation for  the
probability $P_t ({\cal C} ) $ to be in  configuration ${\cal C}$
 at time t
\begin{eqnarray}
\frac{ dP_t \left({\cal C} \right) }{dt}
= \sum_{\cal C '} P_t \left({\cal C}' \right) 
W \left({\cal C}' \to  {\cal C}  \right) 
 -  P_t \left({\cal C} \right) 
\left[  \sum_{ {\cal C} '} W \left({\cal C} \to  {\cal C}' \right) \right]
\label{master}
\end{eqnarray}
where the transition rates $W$ satisfy the detailed balance property
\begin{eqnarray}
e^{- \beta U({\cal C})}   W \left( \cal C \to \cal C '  \right)
= e^{- \beta U({\cal C '})}   W \left( \cal C' \to \cal C   \right)
\label{detailed}
\end{eqnarray}

In the following, we focus on the Metropolis single-spin-flip dynamics :
 the configuration $ {\cal C}=(S_1,S_2,...,S_L) $
containing $L$ spins is connected 
only to the $L$ configurations $ {\cal C}_k=(S_1,S_2,.,-S_k,..,S_L) $ 
obtained by the flip of the single spin $S_k \to -S_k$
with the Metropolis rate
\begin{eqnarray}
W \left( {\cal C} \to {\cal C}_k \right)
= \frac{1}{\tau_0 } {\rm min} \left[ 1,   e^{- \beta \left[ U({\cal C }_k)-U({\cal C }) \right] } \right] 
\label{Wmetropolis}
\end{eqnarray}
where $\tau_0$ is the characteristic time to attempt a spin-flip :
so a spin-flip that decreases the energy $\Delta U=U({\cal C }_k)-U({\cal C }) <0$
has the rate $ \frac{1}{\tau_0 }$, whereas a spin-flip that increases the energy
 $\Delta U=U({\cal C }_k)-U({\cal C }) >0$,  has the rate $ \frac{1}{\tau_0 }e^{-\beta \Delta U}$.

\subsection{ Relaxation spectrum }

As is well known (see for instance the textbooks \cite{gardiner,vankampen,risken}),
 the non-symmetric master Eq.  \ref{master}
can be transformed via the change of variable
\begin{eqnarray}
P_t ( {\cal C} ) \equiv e^{-  \frac{\beta}{2} U(\cal C ) } \psi_t ({\cal C} )
=   e^{-  \frac{\beta}{2} U(\cal C ) } <{\cal C} \vert  \psi_t  >
\label{relationPpsi}
\end{eqnarray}
into the imaginary-time  Schr\"odinger equation
for the ket  $\vert  \psi_t  >$ 
\begin{eqnarray}
\frac{ d }{dt} \vert  \psi_t  > = -H \vert  \psi_t  > 
\label{Hquantum}
\end{eqnarray}
where the quantum Hamiltonian 
\begin{eqnarray}
{\cal H} = \sum_{\cal C } \epsilon \left( {\cal C} \right) \vert {\cal C} > < {\cal C } \vert
+ \sum_{{\cal C},{\cal C '}}  V({\cal C} , {\cal C '})
 \vert {\cal C '} > < {\cal C } \vert
\label{tight}
\end{eqnarray}
contains the symmetric hoppings (Eq \ref{Wmetropolis})
\begin{eqnarray}
 V({\cal C } \to {\cal C }_k) && = - \sqrt {W \left( {\cal C} \to {\cal C }_k \right) W \left( {\cal C }_k \to {\cal C} \right) } 
\nonumber \\
&& = - \frac{1}{\tau_0 }   e^{- \frac{\beta}{2} \vert U({\cal C }_k)-U({\cal C }) \vert } 
= V({\cal C }_k \to {\cal C })
\label{hopping}
\end{eqnarray}
and the on-site energies 
\begin{eqnarray}
 \epsilon \left( {\cal C} \right)
= \sum_{ {\cal C}_k} W \left({\cal C} \to  {\cal C}_k \right)
= \sum_{ {\cal C}_k} \frac{1}{\tau_0 } {\rm min} \left[ 1,   e^{- \beta \left[ U({\cal C }_k)-U({\cal C }) \right] } \right] 
\label{eps}
\end{eqnarray}

In terms of the eigenvalues $E_n$ and 
the associated normalized eigenvectors $\vert \psi_n>$
 of the quantum Hamiltonian  ${\cal H}$ 
\begin{eqnarray}
{\cal H}  \vert  \psi_n > && = E_n \vert \psi_n> \\
\sum_{\cal C} \vert \psi_n({\cal C}) \vert ^2 && =1
\label{spectreH}
\end{eqnarray}
the evolution operator $e^{-t H}$ can be expanded as
\begin{eqnarray}
  e^{-t \cal H}  = 
 \sum_n e^{- E_n t} \vert \psi_n> < \psi_n\vert
\label{spectreHexp}
\end{eqnarray}
so that
the probability 
$P_t \left( {\cal C} \vert {\cal C}_0\right)$ to be in configuration ${\cal C}$ at $t$
if one starts from the configuration ${\cal C}_0$ at time $t=0$
reads
\begin{eqnarray}
P_t \left( {\cal C} \vert {\cal C}_0\right) =
   e^{-  \frac{\beta}{2} \left[U({\cal C} )-U({\cal C}_0 ) \right] } <{\cal C} \vert e^{-t\cal  H}  \vert {\cal C}_0>
= e^{-  \frac{\beta}{2} \left[U({\cal C} )-U({\cal C}_0 ) \right] }
\sum_n e^{- E_n t} \psi_n({\cal C})\psi_n^*({\cal C}_0)
\label{expansionP}
\end{eqnarray}

The quantum Hamiltonian $\cal H$ has the following well-known properties

(i) the ground state energy is $E_0=0$, and the corresponding
eigenvector reads

\begin{eqnarray}
\vert  \psi_0 > = \sum_{\cal C}  \frac{ e^{- \frac{\beta}{2} U({\cal C}) }}{\sqrt Z}
\vert { \cal C}  >
\label{psi0}
\end{eqnarray}
the normalization $1/\sqrt Z$ coming from the quantum normalization of Eq. \ref{spectreH}.
This property ensures the convergence towards the Boltzmann equilibrium
 in Eq. \ref{relationPpsi} for any initial condition ${\cal C}_0$
\begin{eqnarray}
P_t \left( {\cal C} \vert {\cal C}_0\right)
\opsimeq_{t \to + \infty}  e^{-  \frac{\beta}{2} \left[U({\cal C} )-U({\cal C}_0 ) \right]}
\psi_0({\cal C})\psi_0^*({\cal C}_0) = \frac{e^{- \beta U({\cal C})}}{Z} = P_{eq}({\cal C})
\label{CVeqP}
\end{eqnarray}

(ii) the other $({\cal N}-1)=(2^L-1)$ (Eq \ref{calN}) 
energies $E_n>0$ determine the relaxation towards equilibrium.
In particular, the lowest non-vanishing energy $E_1$
determines the largest relaxation time of the system 
\begin{eqnarray}
P_t \left( {\cal C} \vert {\cal C}_0\right) - P_{eq}({\cal C})
\opsimeq_{t \to + \infty} e^{- E_1 t}  e^{-  \frac{\beta}{2} \left[U({\cal C} )-U({\cal C}_0) \right] }
\psi_1({\cal C})\psi_1^*({\cal C}_0) 
\label{CVeqP1}
\end{eqnarray}
This property allows to compute this largest relaxation time $1/E_1$  without simulating the dynamics by any  method able
to compute the first excited energy $E_1$ of the quantum Hamiltonian $\cal H$
 \cite{us_quantum,us_dynamic,us_conjugate}.
In this paper, our goal is to construct the whole hierarchy of relaxation times via 
an appropriate real-space renormalization of the dynamics, but before we need a brief reminder 
on the real-space renormalization for the statics.

\section{ Reminder on the renormalization
 to construct the ground states \cite{c_rgsg} }

\label{sec_rgstatic}

As explained in detail in \cite{c_rgsg}, a very simple real-space renormalization
using blocks of two spins can be used for the energy function 
of the Long-Ranged spin-glass of Eq. \ref{sgLR}
or of its Dyson analog of Eq. \ref{sgDyson}.
The idea is that the two ground states
of the internal energy of each block
\begin{eqnarray}
U^{int}_{2i-1,2i} = -J_{2i-1,2i} S_{2i-1} S_{2i}
\label{eint}
\end{eqnarray}
can be parametrized by the renormalized spin
\begin{eqnarray}
(S^R_{2i}=+) && = (S_{2i-1}={\rm sgn} J_{2i-1,2i},S_{2i}=+ )
\nonumber \\
(S^R_{2i}=-) && = (S_{2i-1}=-{\rm sgn} J_{2i-1,2i},S_{2i}=- )
\label{spinrenormi}
\end{eqnarray}

The renormalized coupling between two renormalized spins $(S^R_{2i},S^R_{2j})$ reads
\begin{eqnarray}
J_{2i,2 j}^{(1)} && = J_{2i,2 m} + {\rm sgn} (J_{2i-1, 2i}){\rm sgn} (J_{2j-1, 2j}) J_{2i-1,2 m-1}
\nonumber \\
&&  +{\rm sgn} (J_{2i-1, 2i}) J_{2i-1,2 j}
+ {\rm sgn} (J_{2j-1, 2j}) J_{2i,2 j-1}
\label{rgjfirst}
\end{eqnarray}
The Gaussian distribution of Eq. \ref{gaussian} or the symmetric L\'evy stable laws of Eq. \ref{levy}
are stable for this renormalization rule, so that one only needs to follow the renormalization
of the characteristic scale $\Delta$ (see \cite{c_rgsg} for more details).

For the Dyson model, this leads to the very simple result that the characteristic
scale $\Delta_n^{(p)}$ of the couplings of generation $n$ after $p \leq n $ renormalization steps
 reads in terms of the initial scale $\Delta_n$ of Eq. \ref{deltandysonsg}
 \begin{eqnarray}
\Delta^{(p)}_n && = \left( 4^{\frac{1}{\mu}} \right)^p \Delta_n
 = L_p^{\theta_{\mu}(\sigma)} \left( \frac{ L_p }{L_n } \right)^{\sigma} 
\label{deltarpdyson}
\end{eqnarray}
in terms of the associated length $L_p =2^p$ and $L_n=2^n$,
and of the droplet exponent
\begin{eqnarray}
\theta_{\mu}(\sigma) && = \frac{2}{\mu}-\sigma 
\label{thetasgmusigma}
\end{eqnarray}
of Eq \ref{regionmutheta}.
For the Long-ranged model of Eq. \ref{sgLR}, the renormalization is 
somewhat heavier to write upon iteration, but yields the same droplet exponent
of Eq. \ref{thetasgmusigma} in the region where it is positive 
(Eq \ref{regionmutheta}) on which we focus here.
We refer to \cite{c_rgsg} for more details and consequences on the distribution of the ground state energy, and we turn to the dynamics.

\section{ Elementary renormalization step for the Metropolis dynamics }

\label{sec_rgrule}

In this section, we derive the basic renormalization rule  for the Metropolis single-spin-flip dynamics
in the limit of small temperature.

\subsection{ Master equation associated to a block of two spins }

The elementary renormalization step concerns a block of two spins, say $(S_1,S_2)$,
with the internal energy
\begin{eqnarray}
U^{int}_{1,2}(S_1,S_2) = -J_{1,2} S_{1} S_{2}
\label{uint}
\end{eqnarray}
and with the following Master equation between the four possibles configurations
$(S_1=\pm,S_2=\pm)$
\begin{eqnarray}
\frac{ dP_t \left( S_1,S_2 \right) }{dt}
&& = P_t(-S_1,S_2) W \left( (-S_1,S_2) \to  (S_1,S_2) \right) 
\nonumber \\
&& + P_t(S_1,-S_2) W \left( (S_1,-S_2) \to  (S_1,S_2) \right) 
\nonumber \\
&& -  P_t \left( S_1,S_2  \right)
 \left[  W \left( (S_1,S_2) \to  (-S_1,S_2) \right) 
+  W \left( (S_1,S_2) \to  (S_1,-S_2) \right) \right]
\label{master4}
\end{eqnarray}

To iterate the RG procedure, we will need to consider the slightly generalized
Metropolis dynamics of Eq. \ref{Wmetropolis}, where each spin $S_k$
has its own characteristic time $\tau_k$ to attempt a spin-flip
\begin{eqnarray}
W \left( {\cal C} \to {\cal C}_k \right)
= \frac{1}{\tau_k } {\rm min} \left[ 1,   e^{- \beta \left[ U({\cal C }_k)-U({\cal C }) \right] } \right] 
\label{Wmetropolisgene}
\end{eqnarray}

Besides the two ground-states parametrized by the renormalized spin
(Eq. \ref{spinrenormi})
\begin{eqnarray}
(S_R=+) && = (S_1={\rm sgn} J_{12},S_2=+ )
\nonumber \\
(S_R=-) && = (S_1=-{\rm sgn} J_{12},-S_2=- )
\label{spinrenorm}
\end{eqnarray}
we need here to introduce also the two excited states 
\begin{eqnarray}
(E=+) && = (S_1=-{\rm sgn} J_{12},S_2=+ )
\nonumber \\
(E=-) && = (S_1={\rm sgn} J_{12},-S_2=- )
\label{spinexc}
\end{eqnarray}

The energies of these configurations are (Eq. \ref{uint})
\begin{eqnarray}
U(S_R=+) && = U(S_R=-)  = -\vert  J_{12} \vert
\nonumber \\
U(E=+) && = U(E=-)  = + \vert  J_{12} \vert
\label{Uspinrenorm}
\end{eqnarray}
whereas the Metropolis transition rates between them read (Eq. \ref{Wmetropolisgene})
\begin{eqnarray}
W \left((S_R=+) \to (E=+) \right) && = \frac{1}{\tau_1 } e^{- 2 \beta \vert  J_{12} \vert }
\nonumber \\
W \left((S_R=+) \to (E=-) \right) && = \frac{1}{\tau_2 } e^{- 2 \beta \vert  J_{12} \vert }
\nonumber \\
W \left((S_R=-) \to (E=+) \right) && = \frac{1}{\tau_2 } e^{- 2 \beta \vert  J_{12} \vert } 
\nonumber \\
W \left((S_R=-) \to (E=-) \right) && = \frac{1}{\tau_1 } e^{- 2 \beta \vert  J_{12} \vert } 
\nonumber \\
W \left((E=+) \to (S_R=+) \right) && = \frac{1}{\tau_1 } 
\nonumber \\
W \left( (E=+) \to (S_R=-) \right) && = \frac{1}{\tau_2 }
\nonumber \\
W \left((E=-) \to (S_R=+) \right) && = \frac{1}{\tau_2 } 
\nonumber \\
W \left( (E=-) \to (S_R=-) \right) && = \frac{1}{\tau_1 }
\label{Wmetropolisgene4}
\end{eqnarray}
So that the Master equation of Eq. \ref{master4}  now reads
\begin{eqnarray}
\frac{ dP_t (S_R=+) }{dt} && = 
- \left( \frac{1}{\tau_1 }+\frac{1}{\tau_2 } \right)
e^{- 2 \beta \vert  J_{12} \vert } P_t (S_R=+)
 + \frac{1}{\tau_1 } P_t(E=+) 
 + \frac{1}{\tau_2 }P_t(E=-)
\nonumber \\
\frac{ dP_t (S_R=-) }{dt} && = 
- \left( \frac{1}{\tau_1 }+\frac{1}{\tau_2 } \right)
e^{- 2 \beta \vert  J_{12} \vert } P_t (S_R=-)
 + \frac{1}{\tau_2 } P_t(E=+) 
 + \frac{1}{\tau_1 }P_t(E=-)
\nonumber \\
\frac{ dP_t (E=+) }{dt} && = 
- \left( \frac{1}{\tau_1 }+\frac{1}{\tau_2 } \right)
 P_t (E=+)
 + \frac{1}{\tau_1 } e^{- 2 \beta \vert  J_{12} \vert } P_t(S_R=+) 
 + \frac{1}{\tau_2 } e^{- 2 \beta \vert  J_{12} \vert } P_t(S_R=-)
\nonumber \\
\frac{ dP_t (E=-) }{dt} && = 
- \left( \frac{1}{\tau_1 }+\frac{1}{\tau_2 } \right)
 P_t (E=-)
 + \frac{1}{\tau_2 } e^{- 2 \beta \vert  J_{12} \vert } P_t(S_R=+) 
 + \frac{1}{\tau_1 } e^{- 2 \beta \vert  J_{12} \vert } P_t(S_R=-)
\label{master4expli}
\end{eqnarray}

\subsection{ Renormalization of the Master equation   }

The renormalization of master equations
containing rapid and slow modes has for goal the elimination of the rapid modes
in order to obtain an effective dynamics for the slow modes.
The explicit renormalization rules for arbitrary master equations 
have been discussed in detail in  \cite{pigolotti,us_rgconfig}.
So here  we simply describe directly how it can be applied to the specific case of Eq.
\ref{master4expli}.

In the limit of small temperature $\beta \vert J_{12} \vert \gg 1$, the Arrhenius factor $e^{\beta \vert J_{12} \vert}$
is huge so that
the effective slow dynamics between the two ground-states configurations $S_R=\pm$
can be found by eliminating the excited states $E=\pm$ that rapidly disintegrate.
We may thus consider
 that these two excited states are in quasi-equilibrium with respect to the slow modes,
so in practice we may set
 $\frac{ dP_t (E=+) }{dt} \simeq 0$ and $\frac{ dP_t (E=+) }{dt} \simeq 0 $
in the two last equations of the system (Eq. \ref{master4expli}) that yield
\begin{eqnarray}
 P_t (E=+)
 \simeq  e^{- 2 \beta \vert  J_{12} \vert }
\ \ \frac{ \frac{1}{\tau_1 } P_t(S_R=+) 
 + \frac{1}{\tau_2 }  P_t(S_R=-) }
{\frac{1}{\tau_1 }+\frac{1}{\tau_2 }}
\nonumber \\
 P_t (E=-) \simeq e^{- 2 \beta \vert  J_{12} \vert }
\ \  \frac{ \frac{1}{\tau_2 }P_t(S_R=+) 
  \frac{1}{\tau_1 }  P_t(S_R=-) }{\frac{1}{\tau_1 }+\frac{1}{\tau_2 }}
\label{master4adiab}
\end{eqnarray}
Plugging these expressions into the two first equations of the system 
(Eq. \ref{master4expli})
yields the effective master equation for the slow modes
\begin{eqnarray}
\frac{ dP_t (S_R=+) }{dt} && = 
- \frac{1}{\tau_R } P_t (S_R=+)
 + \frac{1}{\tau_R } P_t(S_R=-) 
\nonumber \\
\frac{ dP_t (S_R=-) }{dt} && = 
 - \frac{1}{\tau_R } P_t (S_R=-)
 + \frac{1}{\tau_R } P_t(S_R=+) 
\label{master4eff}
\end{eqnarray}
with the following renormalized flip-rate for the renormalized spin $S_R$
\begin{eqnarray}
\tau_R =e^{ 2 \beta \vert  J_{12} \vert } \frac{\tau_1+\tau_2}{2}
\label{rgtauR}
\end{eqnarray}

\section{ Full hierarchy of relaxation times in each disordered sample }

\label{sec_dynamics}

In this section, we explain how the basic renormalization rule of Eq. \ref{rgtauR} for the Metropolis dynamics
in the limit of small temperature
can be used to construct explicitly the full hierarchy of relaxation times in each given disordered sample.
For notational convenience, we have chosen to use here the Dyson notation for the couplings (Eq \ref{sgDyson}),
but the case of the long-ranged model of Eq. \ref{sgLR} can be studied along the same lines.

\subsection{ First RG step : Relaxation towards local equilibrium on each of the $\frac{L}{2}$ blocks of two spins  }

For the initial system of $L=2^N$ spins $S_i$,
we first consider separately the $\frac{L}{2}=2^{N-1}$
blocks of two spins $(S_{2i-1},S_{2i})$ linked by couplings of the zero generation $J_0(2i-1,2i)$.

(i) on the time-scale $\tau_0$, the two excited states of the block (Eq. \ref{spinexc2})
\begin{eqnarray}
(E_{2i}^{R1}=+) && = (S_{2i-1}=-{\rm sgn} J_0(2i-1,2i),S_{2i}=+ )
\nonumber \\
(E_{2i}^{R1}=-) && = (S_{2i-1}={\rm sgn} J_0(2i-1,2i),S_{2i}=- )
\label{spinexc2}
\end{eqnarray}
disintegrate towards one of the two ground states of the block parametrized by the renormalized spin
\begin{eqnarray}
(S^{R1}_{2i}=+) && = (S_{2i-1}={\rm sgn} J_0(2i-1,2i),S_{2i}=+ )
\nonumber \\
(S^{R1}_{2i}=-) && = (S_{2i-1}=-{\rm sgn} J_0(2i-1,2i),S_{2i}=- )
\label{spinrenormi2}
\end{eqnarray}

(ii) The convergence towards the local equilibrium between the two ground states of Eq. \ref{spinrenormi2}
is characterized by the renormalized effective flipping time for the renormalized spin $S^{R1}_{2i}$ (Eq. \ref{rgtauR}
 for the special case $\tau_1=\tau_2=\tau_0$)
\begin{eqnarray}
\tau_{S^{R1}_{2i}} = \tau_0 e^{ 2 \beta \vert  J_0(2i-1,2i) \vert } 
\label{rgtauRfirst}
\end{eqnarray}

For the future dynamics, we may now completely forget the excited states of Eq. \ref{spinexc2},
and keep only the renormalized spins of Eq. \ref{spinrenormi2}, that are characterized by their 
disorder-dependent flipping times of Eq. \ref{rgtauRfirst}.
Two renormalized spins $(S^{R1}_{2i},S^{R1}_{2j})$ are now coupled by the 
 renormalized couplings of Eq \ref{rgjfirst} for $n \geq 1$
\begin{eqnarray}
J_n^{(R1)}(2i,2j) && = J_{n}(2i,2j)
 + {\rm sgn} [J_0(2i-1, 2i){\rm sgn} (J_{0}(2j-1, 2j)] J_n(2i-1,2j-1)
\nonumber \\
&&  +{\rm sgn} [J_0(2i-1, 2i)] J_n(2i-1,2j)
+ {\rm sgn} [J_0(2j-1, 2j)] J_n(2i,2j-1)
\label{rgjfirst2}
\end{eqnarray}

\subsection{ Second RG step : Relaxation towards local equilibrium on each of the $\frac{L}{4}$ blocks of four spins   }

We now consider separately the $\frac{L}{4}=2^{N-2}$ blocks of four initial spins $(S_{4i-3},S_{4i-2},S_{4i-1},S_{4i})$
i.e. of two renormalized spins $(S^{R1}_{4i-2},S^{R1}_{4i})$ linked by renormalized couplings of the  
first generation $J_1^{R1}(2i-2,2i)$ that read in terms of the original couplings (Eq. \ref{rgjfirst2})
\begin{eqnarray}
J_1^{(R1)}(4i-2,4i) && = J_{1}(4i-2,4i)
 + {\rm sgn} [J_0(4i-3, 4i-2){\rm sgn} (J_{0}(4i-1, 4i)] J_1(4i-3,4i-1)
\nonumber \\
&&  +{\rm sgn} [J_0(4i-3, 4i-2)] J_1(4i-3,4i)
+ {\rm sgn} [J_0(4i-1, 4i)] J_1(4i-2,4i-1)
\label{rgjfirst11}
\end{eqnarray}

The convergence towards the local equilibrium between the two corresponding ground states 
parametrized by the renormalized spin of the second generation $R2$
\begin{eqnarray}
(S^{R2}_{4i}=+) && = (S^{R1}_{4i-2}={\rm sgn} J_1^{(R1)}(4i-2,4i),S^{R1}_{4i}=+ )
\nonumber \\
(S^{R2}_{4i}=-) && =  (S^{R1}_{4i-2}=- {\rm sgn} J_1^{(R1)}(4i-2,4i),-S^{R1}_{4i}=- )
\label{spinrenormi4}
\end{eqnarray}
after the elimination of the two excited states of the second generation $R2$
\begin{eqnarray}
(E_{4i}^{R2}=+) && = (S^{R1}_{4i-2}=- {\rm sgn} J_1^{(R1)}(4i-2,4i),S^{R1}_{4i}=+ )
\nonumber \\
(E_{4i}^{R2}=-) && =(S^{R1}_{4i-2}={\rm sgn} J_1^{(R1)}(4i-2,4i),S^{R1}_{4i}=- )
\label{spinexc4}
\end{eqnarray}
is governed by the effective renormalized flipping time (Eq. \ref{rgtauR} and Eq. \ref{rgtauRfirst})
\begin{eqnarray}
\tau_{S^{R2}_{4i}}&&  = e^{ 2 \beta \vert  J_1^{(R1)}(4i-2,4i)) \vert }
 \frac{\tau_{S^{R1}_{4i-2}}+\tau_{S^{R1}_{4i}}}{2}
\nonumber \\
&& = \frac{\tau_0}{2}  e^{ 2 \beta \vert  J_1^{(1)}(4i-2,4i)) \vert }
\left[e^{ 2 \beta \vert  J_0(4i-3,4i-2) \vert }+ e^{ 2 \beta \vert  J_0(4i-1,4i) \vert }  \right]
\label{rgtauRsecond}
\end{eqnarray}

Two renormalized spins $(S^{R2}_{4i},S^{R2}_{4j})$ of the second generation are now coupled by the 
 renormalized couplings of Eq \ref{rgjfirst} for $n \geq 2$
\begin{eqnarray}
J_n^{(R2)}(4i,4j) && = J^{(R1)}_{n}(4i,4j)
 + {\rm sgn} [J^{(R1)}_1(4i-2, 4i){\rm sgn} (J^{(R1)}_{1}(4j-2, 4j)] J^{(R1)}_n(4i-2,4j-2)
\nonumber \\
&&  +{\rm sgn} [J^{(R1)}_1(4i-2, 4i)] J^{(R1)}_n(4i-2,4j)
+ {\rm sgn} [J^{(R1)}_1(4j-2, 4j)] J^{(R1)}_n(4i,4j-2)
\label{rgjfirst4}
\end{eqnarray}

\subsection{ Last RG step : Relaxation towards the equilibrium for the whole sample of $L=2^N$ spins   }

It is now clear how the renormalization procedure has to be iterated up to the last $N$-th RG step,
where the two ground states of the full sample are parametrized by the single renormalized spin of generation $RN$
\begin{eqnarray}
(S^{RN}_{2^N}=+) && = (S^{R(N-1)}_{2^{N-1}}={\rm sgn} J_{N-1}^{(R(N-1))}(2^{N-1},2^N),S^{R(N-1)}_{2^N}=+ )
\nonumber \\
(S^{RN}_{2^N}=-) && =  (S^{R1}_{2^{N-1}}=- {\rm sgn} J_{N-1}^{(R(N-1))}(2^{N-1},2^N), S^{R(N-1)}_{2^N}=- )
\label{spinrenormilast}
\end{eqnarray}
The convergence towards the equilibrium of the whole sample is then governed by 
the effective renormalized flipping time (Eq. \ref{rgtauR})
\begin{eqnarray}
\tau_{S^{RN}_{2^N}}&&  = e^{ 2 \beta \vert  J_{N-1}^{(R(N-1))}(2^{N-1},2^N)) \vert }
 \frac{\tau_{S^{R(N-1)}_{2^{N-1}}}+\tau_{S^{R(N-1)}_{2^N}}}{2}
\label{rgtauRlast}
\end{eqnarray}

Note that the recurrence of Eq. \ref{rgtauRlast} is somewhat similar to the recursions concerning
the partition functions of Derrida's Generalized Random Energy Models \cite{grem} or of the Dyson hierarchical Random Energy Model \cite{dyson_rem}.

The contribution of the last renormalized coupling $ \vert  J_{N-1}^{(R(N-1))}(2^{N-1},2^N)) \vert$ in Eq. \ref{rgtauRlast}
grows with respect to the length $L_{N-1}=2^{N-1}$ with the droplet exponent $\theta_{\mu}(\sigma)$ (Eq. \ref {deltarpdyson})
\begin{eqnarray}
\vert  J_{N-1}^{(R(N-1))}(2^{N-1},2^N)) \vert \propto \Delta^{(N-1)}_{N-1} && = L_{N-1}^{\theta_{\mu}(\sigma)} 
\label{deltarpdysonlast}
\end{eqnarray}
so that the dynamical exponent $\psi$ introduced in Eqs \ref{defpsi} and \ref{ttyp} satisfies the usual bound \cite{Fis_Hus}
\begin{eqnarray}
\psi \geq \theta
\label{psitheta}
\end{eqnarray}
as it should.
To better understand the contributions of all other generations on the final relaxation time,
we analyze the probability distribution of dynamical barriers in the next section.

\section{ Statistics of dynamical barriers }

\label{sec_barrier}

Since the recurrence of Eq. \ref{rgtauRlast} on the relaxation times has been derived near zero-temperature,
it is convenient to focus now on the corresponding dynamical barriers $B(L)$ associated to the various lengths $L$
(Eqs \ref{defpsi} and \ref{ttyp}).

\subsection{ Recurrence on dynamical barriers }

In the present real-space renormalization procedure, the possible lengths are $L_n=2^n$.
For $n=1$ corresponding to the length $L_1=2$, the relaxation times of Eq. \ref{rgtauRfirst}
correspond to the dynamical barriers 
\begin{eqnarray}
B^{R1}_{2i} \equiv \lim_{\beta \to +\infty} \left(  \frac{ \ln \tau_{S^{R1}_{2i}} }{\beta } \right)= 
 2  \vert  J_0(2i-1,2i) \vert  
\label{barrierfirst}
\end{eqnarray}
Since the coupling $J_0(2i-1,2i)=\Delta_0 \epsilon_{2i-1,2i}$ have for characteristic scale $\Delta_0=1$ (Eq. \ref{deltandysonsg})
and since the $ \epsilon_{2i-1,2i}$ are distributed with the L\'evy stable law of index $1<\mu \leq 2$ (Eqs \ref{gaussian} or \ref{levy}), the probability distribution of the barriers $B^{R1}$ reads for $B^{R1} $
\begin{eqnarray}
P_{R1}(B^{R1}) = 2 L_{\mu} \left(  \frac{B^{R1}}{2} \right) 
\label{pbarrierfirst}
\end{eqnarray}

Then the recurrence of Eq. \ref{rgtauRlast} for relaxation times 
yields that the dynamical barriers associated to the length $L_n=2^n$
\begin{eqnarray}
B^{Rn} \equiv \lim_{\beta \to +\infty} \left(  \frac{ \ln \tau_{S^{Rn}} }{\beta } \right)
\label{barrierndef}
\end{eqnarray}
satisfy the recurrence
\begin{eqnarray}
B^{R(n+1)} =  2   \vert  J_n^{(n)}\vert
+  {\rm max} \left[B^{R(n)}_{a} , B^{R(n)}_{b}  \right] 
\label{barriern}
\end{eqnarray}
in terms of two statistically independent barriers $B^{R(n)}_{a}$ and $ {\cal B}^{R(n)}_{b}  $ 
of the generation $R(n)$, and of the 
renormalized coupling $J_n^{(n)}=\Delta_n^{(n)} \epsilon $ with the characteristic scale  (Eq. \ref{deltarpdyson} ) 
\begin{eqnarray}
\Delta_n^{(n)}=2^{n \theta_{\mu}(\sigma)}
\label{deltann}
\end{eqnarray}
in terms of the droplet exponent $ \theta_{\mu}(\sigma)$, 
and where $\epsilon$ is distributed with the L\'evy stable law of index $1<\mu \leq 2$ (Eqs \ref{gaussian} or \ref{levy}).
As a consequence, the probability distributions of dynamical barriers satisfy the recurrence
\begin{eqnarray}
P_{R(n+1)}(B^{R(n+1)}) && = \int_{-\infty}^{+\infty} d \epsilon L_{\mu}(\epsilon) 
\int_0^{+\infty} dB^{R(n)}_{a}   P_{R(n)}(B^{R(n)}_{a}) 
\int_0^{+\infty} dB^{R(n)}_{b}   P_{R(n)}(B^{R(n)}_{b}) 
\nonumber \\
&& \delta \left[  B^{Rn} - \left( 2 \Delta_n^{(n)} \vert \epsilon \vert + {\rm max} \left[B^{R(n)}_{a} , {\cal B}^{R(n)}_{b}  \right] 
\right)
 \right]
\nonumber \\
&& = 4 \int_{0}^{+\infty} d \epsilon L_{\mu}(\epsilon) 
\int_0^{+\infty} dB^{R(n)}_{a}   P_{R(n)}(B^{R(n)}_{a}) 
\int_{0} ^{B^{R(n)}_{a} } dB^{R(n)}_{b}   P_{R(n)}(B^{R(n)}_{b}) 
\nonumber \\
&& \delta \left[  B^{Rn} - \left( 2 \Delta_n^{(n)}  \epsilon  + B^{R(n)}_{a} \right) \right]
\label{barrierpdfiter}
\end{eqnarray}
with the initial condition of Eq. \ref{pbarrierfirst}.
So it is clear that the probability distribution of dynamical barriers
will keep the same type of asymptotic tail as $L_{\mu}$.
It is thus convenient now to discuss separately the Gaussian case $\mu=2$
and the L\'evy cases $1<\mu<2$ displaying power-law tails.

\subsection { Analysis of the barrier distribution for the Gaussian case $\mu=2$ } 

In the Gaussian case $\mu=2$, where the kernel $L_{\mu}$ of Eq. \ref{barrierpdfiter} is Gaussian (Eq. \ref{gaussian}),
Eq. \ref{barrierpdfiter} reads
\begin{eqnarray}
P_{R(n+1)}(B^{R(n+1)}) && = 2 \frac{1}{\sqrt{ \pi} } \int_{0}^{+\infty} d \epsilon  e^{- \frac{\epsilon^2}{4}}
\int_0^{+\infty} dB^{R(n)}_{a}   P_{R(n)}(B^{R(n)}_{a}) 
\int_{0} ^{B^{R(n)}_{a} } dB^{R(n)}_{b}   P_{R(n)}(B^{R(n)}_{b}) 
\nonumber \\
&& \delta \left[  B^{Rn} - \left( 2 \Delta_n^{(n)}  \epsilon  + B^{R(n)}_{a} \right) \right]
\label{barrierpdfitergauss}
\end{eqnarray}
with the Gaussian initial condition (Eq. \ref {pbarrierfirst})
\begin{eqnarray}
P_{R1}(B^{R1}) = 
 \frac{1}{\sqrt{ \pi} } e^{- \frac{1}{4} \left(  \frac{B^{R1}}{2} \right)^2}
\label{barrierfirstgauss}
\end{eqnarray}
As a consequence, the probability distribution of barriers will keep the same type of tail 
\begin{eqnarray}
P_{Rn}(B^{Rn}) \oppropto_{ B^{Rn} \to +\infty}  
 e^{- \frac{1}{4} \left(  \frac{B^{Rn}}{\Gamma_n} \right)^2}
\label{barriertailgauss}
\end{eqnarray}
where $\Gamma_n$ is the appropriate scale governing the decay at infinity.
A saddle point evaluation of Eq. \ref{barrierpdfiter} yields the following recurrence 
for the scale $\Gamma_n$
\begin{eqnarray}
\Gamma_{n+1}^2 = 4 (\Delta_n^{(n)} )^2+\Gamma_{n}^2
\label{itergamman}
\end{eqnarray}
with the initial condition $\Gamma_{n=1}=2 $ (Eq. \ref{barrierfirstgauss}) or equivalently $\Gamma_{n=0}=0 $.
Using Eq. \ref{deltann}, one obtains
\begin{eqnarray}
\Gamma_{n}^2 = 4 \sum_{k=0}^{n-1}  (\Delta_k^{(k)} )^2 = 4 \sum_{k=0}^{n-1} (2^{2  \theta_{2}(\sigma)})^k
= 4 \frac{2^{2 n \theta_{2}(\sigma)} -1 }{2^{2  \theta_{2}(\sigma)}-1}
\label{solgamman}
\end{eqnarray}

For large $n$, we thus obtain that the characteristic scale $\Gamma_n$ that governs the asymptotic decays of
Eq. \ref{barriertailgauss} grows with respect to the length $L_n=2^n$ as
\begin{eqnarray}
\Gamma_{n} \simeq   \frac{ 2 }{\sqrt{ 2^{2  \theta_{2}(\sigma)}-1 } } L_n^{\theta_{2}(\sigma)}
\label{resgammangauss}
\end{eqnarray}
so that the dynamical exponent $\psi$ introduced in Eq. \ref{defpsi} here coincides with the droplet exponent $ \theta_{2}(\sigma)$ (Eq. \ref{thetaLRd1gauss})
\begin{eqnarray}
\psi=\theta_{2}(\sigma) =1-\sigma
\label{psigauss}
\end{eqnarray}
i.e. it saturates the bound $\psi \geq \theta$ of Eq. \ref{psitheta} \cite{Fis_Hus},
whereas in short-ranged Gaussian spin-glasses, one expects the strict inequality
$\psi>\theta$ (see \cite{us_conjecture} and references therein).

\subsection{ Analysis of the barrier distribution for the L\'evy cases $1<\mu<2$ }

In the L\'evy cases $1<\mu<2$, both the kernel $L_{\mu}$ of Eq. \ref{barrierpdfiter} and the initial condition of Eq. \ref{pbarrierfirst} display the 
power-law decay of Eq. \ref{levytail}.
 As a consequence, it is clear that the probability distribution of barriers will also display the same power-law tail
\begin{eqnarray}
P_{Rn}(B^{Rn}) \oppropto_{ B^{Rn} \to +\infty}  
 \frac{A_{\mu}}{ B^{Rn}}  \left( \frac{ \Gamma_n }{ B^{Rn}}  \right)^{\mu}
\label{levytailbarrier}
\end{eqnarray}
where $\Gamma_n$ is the appropriate scale governing the decay at infinity.
The asymptotic analysis of the iteration of Eq. \ref{barrierpdfiter}
yields the recurrence
\begin{eqnarray}
\Gamma_{n+1}^{\mu} =  ( 2\Delta_n^{(n)} )^{\mu}+2 \Gamma_{n}^{\mu}
\label{itergammanlevy}
\end{eqnarray}
with the initial condition $\Gamma_{n=0}=0 $.
Using Eq. \ref{deltann}, one obtains
\begin{eqnarray}
\Gamma_{n}^{\mu} =  \sum_{k=0}^{n-1}  (2 \Delta_k^{(k)} )^{\mu} 2^{ n-1 -k} 
= 2^{\mu+n-1} \sum_{k=0}^{n-1}   \left( 2^{ \mu \theta_{\mu}(\sigma) -1 }  \right)^k
= 2^{\mu+n-1} \frac{  2^{ n \left(\mu \theta_{\mu}(\sigma) -1 \right) }  -1 }{2^{ \mu \theta_{\mu}(\sigma) -1 } -1}
\label{solgammanlevy}
\end{eqnarray}

In the region of parameters that we consider (Eq. \ref{regionmutheta}), it turns out that 
$ \left(\mu \theta_{\mu}(\sigma) -1 \right) = 1 - \mu \sigma <0$,
and the characteristic scale $\Gamma_n$ grows with respect to the length $L_n=2^n$ as
\begin{eqnarray}
\Gamma_{n} \simeq   \frac{ 2^{1-\frac{1}{\mu}}}{ \left( 1- 2^{ \mu \theta_{\mu}(\sigma) -1 } \right)^{\frac{1}{\mu}} }
L_n^{\frac{1}{\mu}}
\label{resgammanlevy}
\end{eqnarray}
so that the dynamical exponent $\psi$ introduced in Eq. \ref{defpsi} here 
does not coincide with the droplet exponent $ \theta_{\mu}(\sigma)=\frac{2}{\mu}-\sigma$ 
but is bigger
\begin{eqnarray}
\psi=\frac{1}{\mu} 
\label{psilevy}
\end{eqnarray}

This value can  
be understood by the simple following scaling argument concerning 
only the largest barrier $B^{R1}_{max}(L)$ among the $L$ barriers $B^{R1}$ of the smallest scale
 (Eq. \ref{barrierfirst}) distributed with a power-law tail
\begin{eqnarray}
\frac{1}{L} \propto \int_{B_{max}(L)}^{+\infty} \frac{dB^{R1}}{(B^{R1})^{1+\mu}} 
\propto \frac{1}{(B^{R1}_{max}(L))^{\mu}}
\label{psilevyarg}
\end{eqnarray}
leading to $B^{R1}_{max}(L) \propto L^{\frac{1}{\mu}} $.
It seems thus interesting to apply the same argument for the short-ranged spin-glass
on the hypercubic lattice of dimension $d$ with L\'evy couplings.
The largest barrier $B^{R1}_{max}(N=L^d)$ among $N=L^d$ barriers of the smallest scale
distributed with a power-law tail scales as  $B^{R1}_{max}(N=L^d) \propto  N^{\frac{1}{\mu}}
=  L^{\frac{d}{\mu}}$ suggesting the following barrier exponent for $1<\mu<2$.
\begin{eqnarray}
\psi_{\mu}^{SR}=\frac{d}{\mu}
\label{psilevySR}
\end{eqnarray}

\section{ Conclusion }

\label{sec_conclusion}

We have considered the long-ranged Ising spin-glass with random couplings decaying as a power-law of the distance, in the region of parameters 
where the energy is extensive, and where the spin-glass phase exists with a positive droplet exponent. For the Metropolis single-spin-flip dynamics near zero temperature, we have constructed  via real-space renormalization the full hierarchy of relaxation times for any given realization of the random couplings. We have then analyzed the probability distribution of dynamical barriers as a function of the spatial scale. 
The present study represents a simple explicit example of the droplet scaling theory, where the convergence towards local equilibrium on larger and larger scales is governed by a strong hierarchy of activated dynamical processes, with valleys within valleys. 

A natural question is whether the present analysis concerning spin-glasses (i) with Long-Ranged interactions (ii) with a positive droplet exponent $\theta>0$, could be generalized to study the dynamics of long-ranged spin-glasses with negative droplet exponents, or of short-ranged spin-glasses with positive or negative droplet exponents.
In our present approach, it is clear that the renormalization of the dynamical barriers
is constructed on the top of a static renormalization near zero-temperature, so the first requirement is to have an appropriate renormalization for the ground-state, which is able to reproduce the correct droplet exponent $\theta$. As explained in \cite{c_rgsg}, the simple block renormalization considered in the present paper reproduces the correct droplet exponent only for Long-Ranged Spin-glasses with a positive droplet exponent $\theta>0$, whereas for Long-Ranged Spin-glasses with negative droplet exponent $\theta<0$, one needs to use a more complicated renormalization procedure where the boundaries of correlated clusters are not fixed a priori but are chosen as a function of the disorder realization (see \cite{c_rgsg} for more details). One expects that the same idea should be used for short-range spin-glasses, but an explicit renormalization procedure to construct the appropriate correlated clusters in each disorder realization is still lacking.

\end{document}